\documentclass[preprintnumbers,showpacs,amsmath,amssymb,floatfix,9pt,prd,onecolumn,
superscriptaddress,nofootinbib]{revtex4}
\usepackage{latexsym}
\usepackage{epsfig}
\usepackage{amssymb}
\begin{document}

\title{\textbf{Conformally symmetric relativistic star}}

 \author{Farook Rahaman}
\email{rahaman@iucaa.ernet.in} \affiliation{Department of
Mathematics, Jadavpur University, Kolkata 700 032, West Bengal,
India}

\author{Sunil D. Maharaj}
\email{maharaj@ukzn.ac.za}
\affiliation{Astrophysics and Cosmology Research Unit, School of Mathematics, Statistics and Computer Science, University of KwaZulu-Natal,Private Bag X54001, Durban  4000, South Africa}

\author{Iftikar Hossain Sardar}
\email{iftikar.spm@gmail.com} \affiliation{ Department of
Mathematics, Jadavpur University, Kolkata 700 032, West Bengal,
India }

\author{Koushik Chakraborty}
\email{koushik@iucaa.ernet.in  } \affiliation{Department of
Physics, Government Training College, Hooghly - 712103, West
Bengal, India}

\date{today}

\begin{abstract}
We investigate whether compact stars having Tolman-like interior geometry admit conformal symmetry. Taking anisotropic pressure along the two principal directions within the compact object, we obtain physically relevant quantities such as transverse and radial pressure, density and redshift function. We study the equation of state for the matter distribution inside the star. From the relation between pressure and density function of the constituent matter, we explore the nature and properties of the interior matter. The red shift function, compactness parameter are found to be physically reasonable. The matter inside the star satisfies the null, weak and strong energy conditions. Finally, we compare the masses and radii predicted from the model with corresponding values in some observed stars.

\end{abstract}

\pacs{04.40.Nr, 04.20.Jb, 04.20.Dw}

\maketitle

\section{Introduction}
~~~In the late stage of stellar evolution when their nuclear fuel gets exhausted, stars turn into highly dense compact objects. The end of the star is decided by the mass of the star. Massive stars explode into supernovae and turn into neutron stars. The general perception is that neutron stars are supported by the neutron degeneracy pressure. It is strongly believed that at such high densities at the core relativistic nucleons must get converted to hyperons, or form condensates. There are predictions that these nucleons might form Cooper pairs and might be in superfluid state. The strange quark matter hypothesis by Witten \cite{witten}, based on MIT Bag Model, indicates that the quarks inside the compact star might not be in a confined hadronic state. At such extreme densities and pressures they might form a larger colorless region with equal proportion of up, down and strange quarks. Thus the constitution of the core region of the compact stars is still an open question for the researchers. Several proposals exist in the literature regarding the EOS of the interior matter \cite{eos, klahn, yan}. In the present investigation we try to attack the problem from an alternative approach. We choose the spherically symmetric spacetime inside the star which is the most probable geometric description for the interior of a compact star, and an assumed conformal symmetry along the radial direction. These prescriptions enable us to avoid the adhoc choice of any EOS for the matter inside the star. Rather, from the context of General Relativity, we should try to infer the nature of the EOS of the constituent matter.

~~~Ruderman \cite{ruderman} pointed out that nuclear matter tends to become anisotropic in nature at very high densities ( $\sim 10^{15} gm/ cc$ ). Ponc\'{e} de Le\'{o}n \cite{ponce} obtained two new exact analytical solutions to Einstein Field Equations for static fluid spheres with anisotropic pressures. Maharaj and Maartens \cite{maharaj3} found new interior solutions for static anisotropic fluid spheres under the assumption of uniform energy density. On the other hand, conformal symmetry in spherical symmetric fluid distributions is an interesting field of study among researchers. Herrera et al. \cite{herrera1} studied under general relativity the nature of the solutions for isotropic and anisotropic fluid spheres admitting one-parameter group of conformal motions. In yet another paper, Herrera and Ponc\'{e} de Le\'{o}n \cite{herrera2}  studied static conformal symmetry in a spherically symmetric charged perfect fluid distribution. They obtained some solutions for isotropic and anisotropic matter. Maartens and Maharaj \cite{maharaj2} found new solutions for static spherical distributions in charged imperfect fluids admitting conformal symmetry. Recently, Rahaman et al. \cite{rahaman1} obtained a new class of interior solutions for an anisotropic star with static conformal symmetry. Ray et al. \cite{ray, rahaman2} studied electromagnetic mass models for static spherical distribution admitting conformal Killing vectors. Esculpi and Alom\'{a} \cite{esculpi} proposed two new solutions for charged spherically symmetric matter distribution admitting conformal symmetry. They assumed a linear equation of state for the matter and anisotropic pressure inside the sphere. In the present paper we consider anisotropic spherically symmetric matter distribution admitting nonstatic conformal symmetry. Unlike Esculpi and Alom\'{a} we do not assume an equation of state for the interior matter; if an equation of state exists it should follow from the field equations.

~~~In the next section we describe the model. In section III the solutions are presented. In section IV the boundary conditions are discussed. We discuss the results in section V. The model is tested under some of the standard stellar criteria in section VI. We finish with some concluding remarks.
\section{The Model}
~~~We take the Tolman-like \cite{tolman} metric to
describe the spacetime inside the compact object. Thus the metric is considered of the following form
\begin{equation}
ds^{2}=-e^{\nu(r)}dt^2+\left(1 + \frac{r^2}{R^2}\right) dr^2+r^2(d\theta^2+\sin^2\theta
d\phi^2), \label{metric}
\end{equation}
where $R$ is the parameter responsible for the geometry of the star.

The  general energy momentum tensor is
\begin{equation}
T_\nu^\mu=  ( \rho + p_r)u^{\mu}u_{\nu} + p_r g^{\mu}_{\nu}+
            (p_t -p_r )\eta^{\mu}\eta_{\nu}, \label{eq:emten}
\end{equation}
with  $ - u^{\mu}u_{\mu} =  \eta^{\mu}\eta_{\mu} = 1$. $u^{\mu}$ is the fluid four velocity .

The Einstein equations for the line element (\ref{metric}) are
\begin{equation}\frac{1}{R^2}\left(3 + \frac{r^2}{R^2}\right)\left(1 + \frac{r^2}{R^2}\right)^{-2} = 8 \pi \rho, \label{eq:lam}
\end{equation}
\begin{equation}\left(1 + \frac{r^2}{R^2}\right)^{-1}\left[\frac{\nu^\prime}{r} + \frac{1}{r^2}\right] - \frac{1}{r^2}=
8\pi p_r, \label{eq:nu}
\end{equation}
\begin{equation}
\left(1 +\frac{r^2}{R^2}\right)^{-1}\left[\frac{\nu^{\prime\prime}}{2} +\frac{\nu^\prime}{2r} +
 \frac{(\nu^{\prime})^{2}}{4}\right]- \frac{1}{R^2}\left(1 + \frac{r^2}{R^2}\right)^{-2}\left[1 + \frac{\nu^\prime r}{2}\right] =8\pi p_t. \label{eq:tan}
 \end{equation}

The study of inheritance symmetry is very useful in searching for the natural relation between geometry
and matter through the Einstein equations. The well known inheritance symmetry is the
symmetry under conformal Killing vectors (CKV). Moopanar and Maharaj \cite{maharaj1} in their paper give an extensive account of the conformal Killing symmetries in spherical spacetimes. There are other exact solutions of the Einstein field equations with the consideration of conformal symmetry \cite{conf, Harko}. Recently, Radinschi et al. \cite{radinschi} proposed nonstatic conformal symmetry for an anisotropic spherical distribution of perfect fluid, to obtain a classical model of electron. Shee et al. \cite{shee} obtained analytical solutions for anisotropic relativistic compact stars admitting nonstatic conformal symmetry.  The non static conformal
Killing equation for (\ref{metric}) becomes
\begin{equation}
L_\xi g_{ij} = g_{ij;k} \xi^{k} + g_{kj} \xi^{k}_{;i}+ g_{ik} \xi^{k}_{;j} = \psi g_{ij}.\label{CKV}
\end{equation}
where $L_\xi$ is the Lie derivative
operator and $\psi$ is the conformal factor. Here the vector $\xi$ generates the conformal symmetry and then
the metric $g_{ik}$ is conformally mapped onto itself along $\xi$.
Herrera et al. \cite{herrera1, herrera2} assumed $\xi$ as nonstatic for a static $\psi$. Following them we can write
\begin{equation}
\xi = \alpha (t,r) \partial_{t} + \beta (t,r) \partial_{r}, \label{xi}
\end{equation}
\begin{equation}
\psi = \psi(r). \label{psi}
\end{equation}
Then equations (\ref{metric}), (\ref{CKV}), (\ref{xi}), (\ref{psi}) can be solved simulataneously to give
\begin{equation}
\alpha = A + \frac{1}{2}Kt,
\end{equation}

\begin{equation}
\beta = \frac{1}{2}Bre^{-\frac{\lambda}{2}},
\end{equation}

\begin{equation}
\psi = Be^{- \frac{\lambda}{2}},
\end{equation}

\begin{equation}
e^{\nu} = C^2 r^2 \exp \left[-2KB^{-1} \int \frac{e^\frac{\lambda}{2}}{r} dr\right] \label{nu}
\end{equation}
where $A, B, C, K$ are constants. Without loss of generality we can choose $A = 0$ and $B = 1$ \cite{maharaj2} for the above solutions and,
\begin{equation}
e^{\lambda} = \left(1 + \frac{r^2}{R^2}\right),
\end{equation}
for consistency with the metric (\ref{metric}).

\section{Solutions}
From the equation (\ref{nu}) we get
\begin{equation}
e^{\nu} = \tilde{C}^2 r^2 \exp \left[-2K \int \frac{\sqrt{\left(1 + \frac{r^2}{R^2}\right)}}{r}dr + D \right].
\end{equation}
This can be simply integrated to give
\begin{equation}
e^{\nu} = \tilde{C}^2 r^2 \exp \left[-2K\left( \sqrt{1 + \frac{r^2}{R^2}} - \tan^{-1}\left(\frac{1}{\sqrt{1 + \frac{r^2}{R^2}}}\right)\right) + D \right],  \label{enu}
\end{equation}
where $D$ is the constant of integration. We can rescale the constant $\tilde{C^2}$ as $ C^2 = \tilde{C^2} e^{D}$.

Then equation (\ref{eq:lam}) gives
\begin{equation}
8\pi \rho = \frac{1}{R^2}\left(3 + \frac{r^2}{R^2}\right)\left(1 + \frac{r^2}{R^2}\right)^{-1}. \label{rho}
\end{equation}
The expression for $p_r$ is
\begin{equation}
8\pi p_r =  \left(1+\frac{r^2}{R^2}\right)^{-1}
 \left[\frac{3}{r^2}-\frac{2K}{\sqrt{1+\frac{r^2}{R^2}}}
\left(\frac{1}{R^2}+\frac{1}{r^2+R^2}\right)\right] -\frac{1}{r^2}.
 \label{pr}
\end{equation}

The expression for $p_t$ is
\begin{equation}
8 \pi p_t = \left(1+\frac{r^2}{R^2}\right)^{-1} \left[ \tilde{A} - \tilde{B} \right] -\frac{1}{R^2}\left(1+\frac{r^2}{R^2}\right)^{-2}\left[2-\frac{Kr^2}{\sqrt{1+\frac{r^2}{R^2}}}\left(\frac{1}{R^2}+\frac{1}{r^2+2R^2}\right) \right],
\end{equation}
where,
$\tilde{A} = \frac{2Kr^2}{\left(r^2+2R^2\right)\sqrt{1+\frac{r^2}{R^2}}} + \left(\frac{1}{r}-\frac{Kr}{\sqrt{1+\frac{r^2}{R^2}}}\left(\frac{1}{R^2} + \frac{1}{r^2+2R^2}\right)\right)^2 $,\\

$\tilde{B} = \frac{K}{\left(1+\frac{r^2}{R^2}\right)^{\frac{3}{2}}}\left(\frac{1}{R^2}+\frac{1}{r^2+R^2}\right) + \frac{K}{\sqrt{1+\frac{r^2}{R^2}}}\left(\frac{1}{R^2}+\frac{1}{r^2+2R^2}\right)$. \\

The mass function is given by
\begin{equation}
m(r) = \int_{0}^{r} 4 \pi r^2 \left( \frac{1}{8 \pi R^2}\left(3 + \frac{r^2}{R^2}\right)\left(1 + \frac{r^2}{R^2}\right)^{-1}\right) dr = \frac{r^3}{600} + r - 10\arctan{\left(\frac{r}{R}\right)}.
\end{equation}

\begin{figure*}
 \begin{tabular}{rl}
 \includegraphics[scale=0.3]{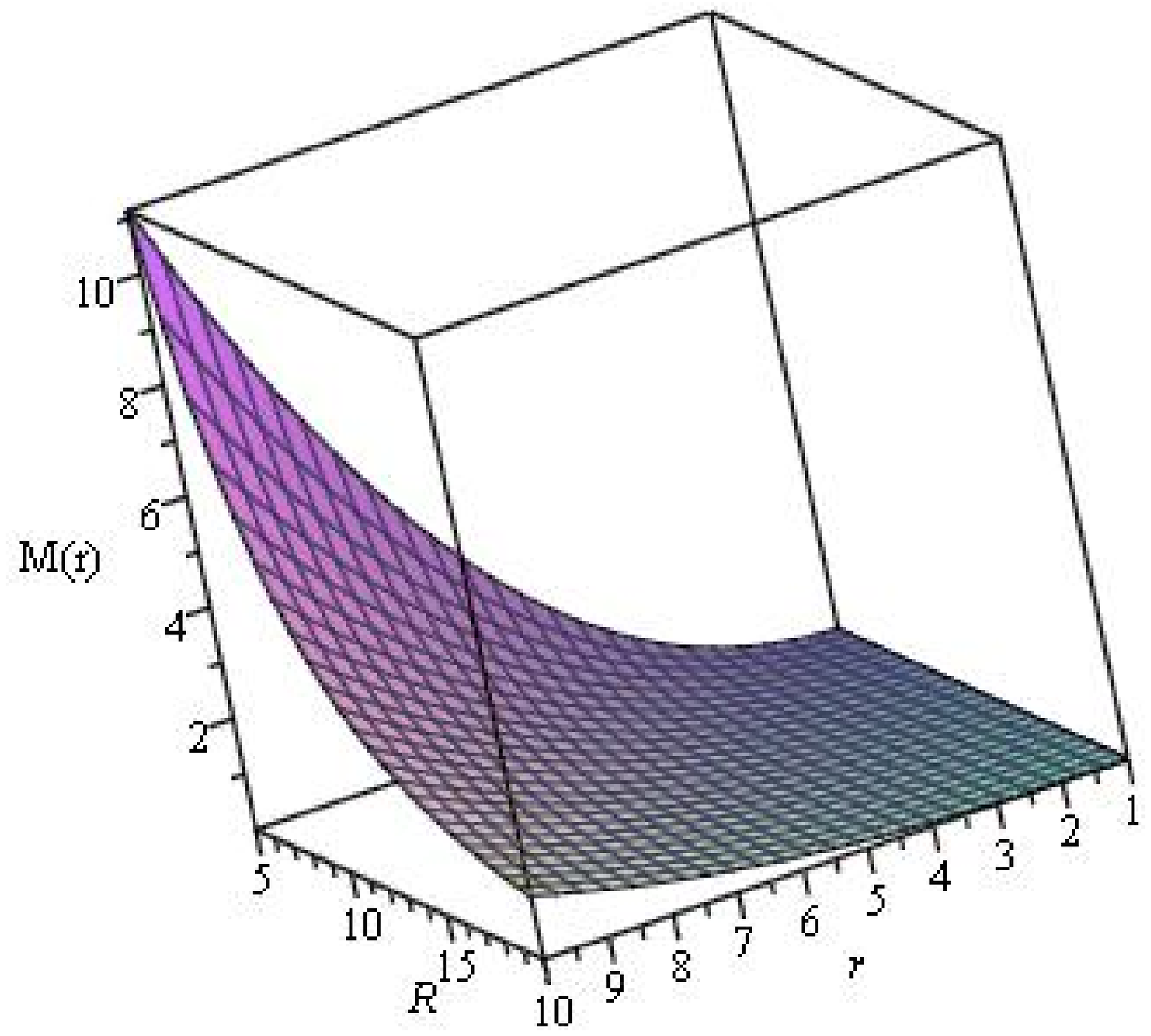}&
 \includegraphics[scale=0.3]{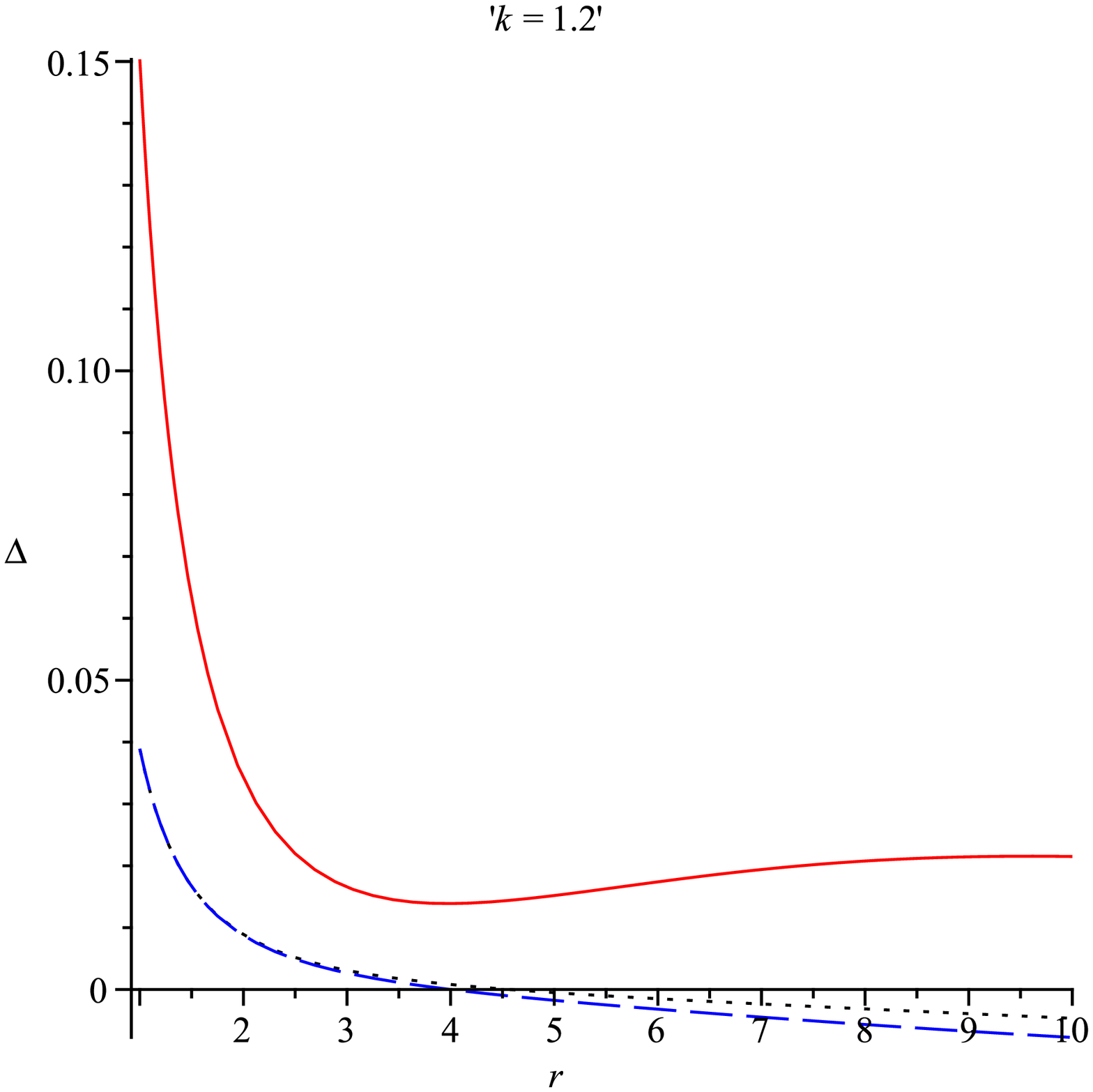}\\
 \end{tabular}
 \caption{\small{(left) Variation of mass function with radial distance $r$ and the scaling papameter $R$. (right) Pressure anisotropy ($\Delta = p_r - p_t$) of the star for $R = 30$ (solid), $R = 20$ (dot) and $R = 10$ (dashed). In geometric units $M(r)$, $r$ and $R$ have unit $km$ and $\Delta$ has unit $km^{-2}$.}}

\end{figure*}

\section{Boundary Conditions}

For a stellar model we must have $p_r ( r = a ) = 0$ where 'a' is the radius of the star:

\begin{equation}
\left(1+\frac{a^2}{R^2}\right)^{-1}
 \left[\frac{3}{a^2}-\frac{2K}{\sqrt{1+\frac{a^2}{R^2}}}
\left(\frac{1}{R^2}+\frac{1}{a^2+R^2}\right)\right] -\frac{1}{a^2} = 0. \label{bc}
\end{equation} \\

Taking the outside geometry of the star to be the Scwarzschild geometry, we get from the boundary condition \\

 $\left(g_{tt}\right)_{r = a} = \left(1 - \frac{2M}{a}\right)$,
\begin{equation}
C^2 a^2 e^{\left(-2 K \sqrt{1 + \frac{a^2}{R^2}}\right)} e^{\left(\arctan\left(\frac{1}{\sqrt{1 + \frac{a^2}{R^2}}}\right)\right)} = \frac{20 \arctan\left(\frac{a}{R}\right)}{a} - 1 - \frac{a^2}{300}, \label{bc1}
\end{equation}

and the condition $\left(g_{rr}\right)_{r = a} = \left(1 - \frac{2M}{a}\right)^{-1}$, gives

\begin{equation}
\frac{20 \left(1 + \frac{a^2}{R^2}\right) \arctan \left(\frac{a}{R} \right)}{a} = \frac{a^4}{300R^2} + a^2 \left( \frac{1}{300} + \frac{1}{R^2} \right) + 2 . \label{bc2}
\end{equation}
 The above equations (\ref{bc}), (\ref{bc1}), (\ref{bc2}) are extremely nonlinear in nature and we can hardly get any exact values for the constants $C^2$, $K$ and $R$. However, from the contour plot of the equation (\ref{bc})  and (\ref{bc2}) we can get an idea of the variation with $a$ and $R$.
\begin{figure*}[thbp]
\begin{tabular}{lr}
 \includegraphics[scale=0.3]{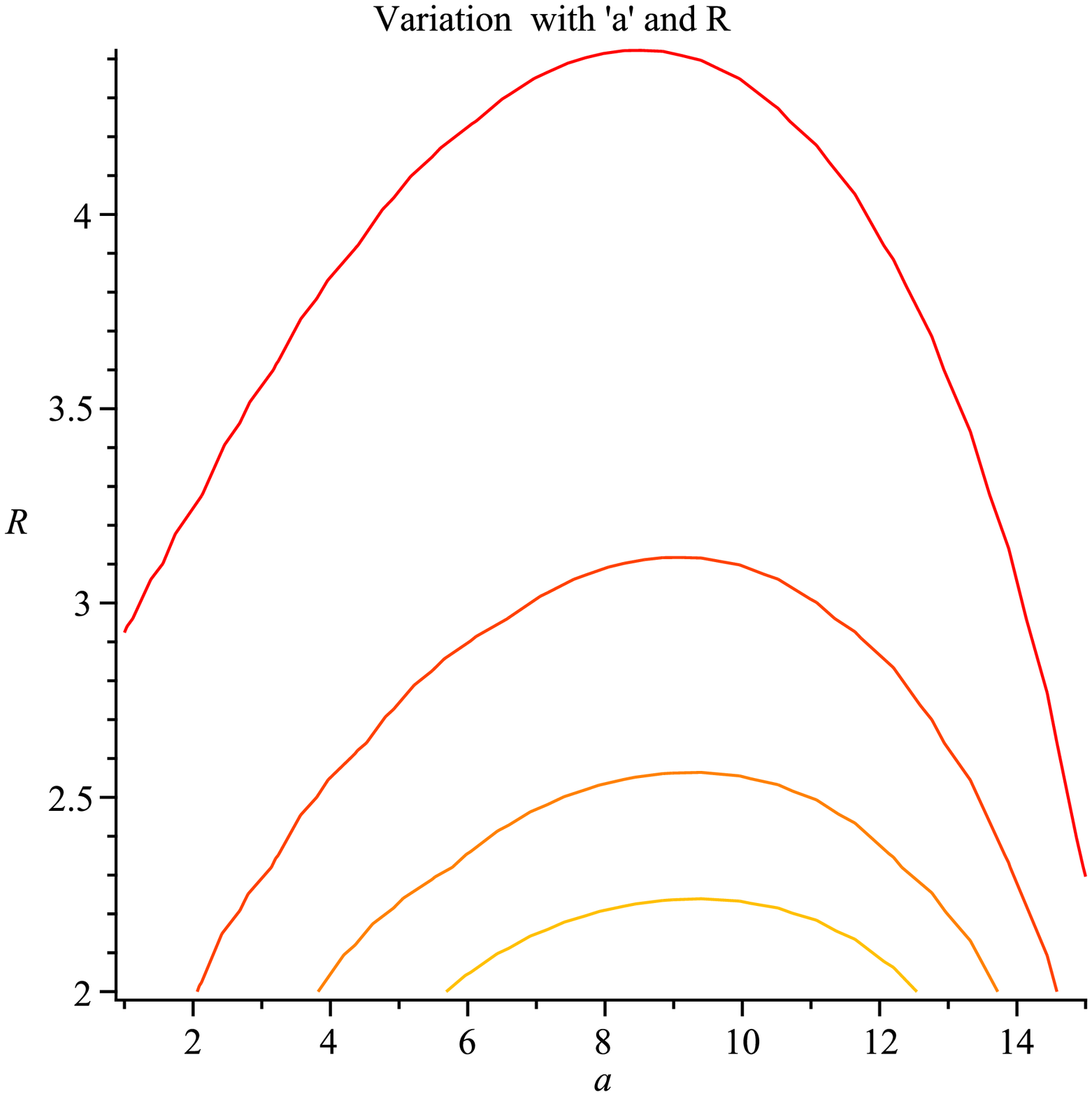}&
 \includegraphics[scale=0.3]{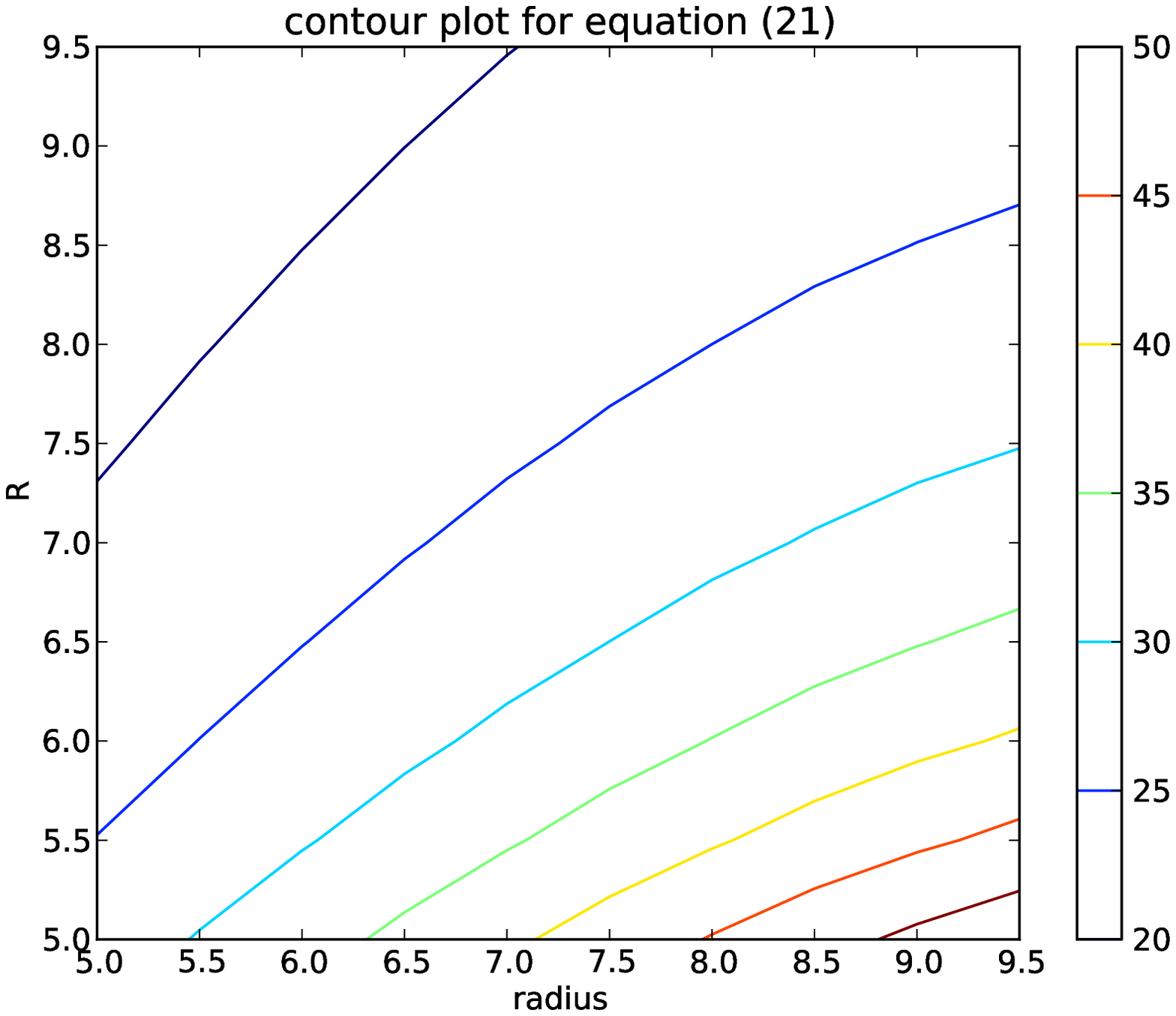}
 \includegraphics[scale=0.3]{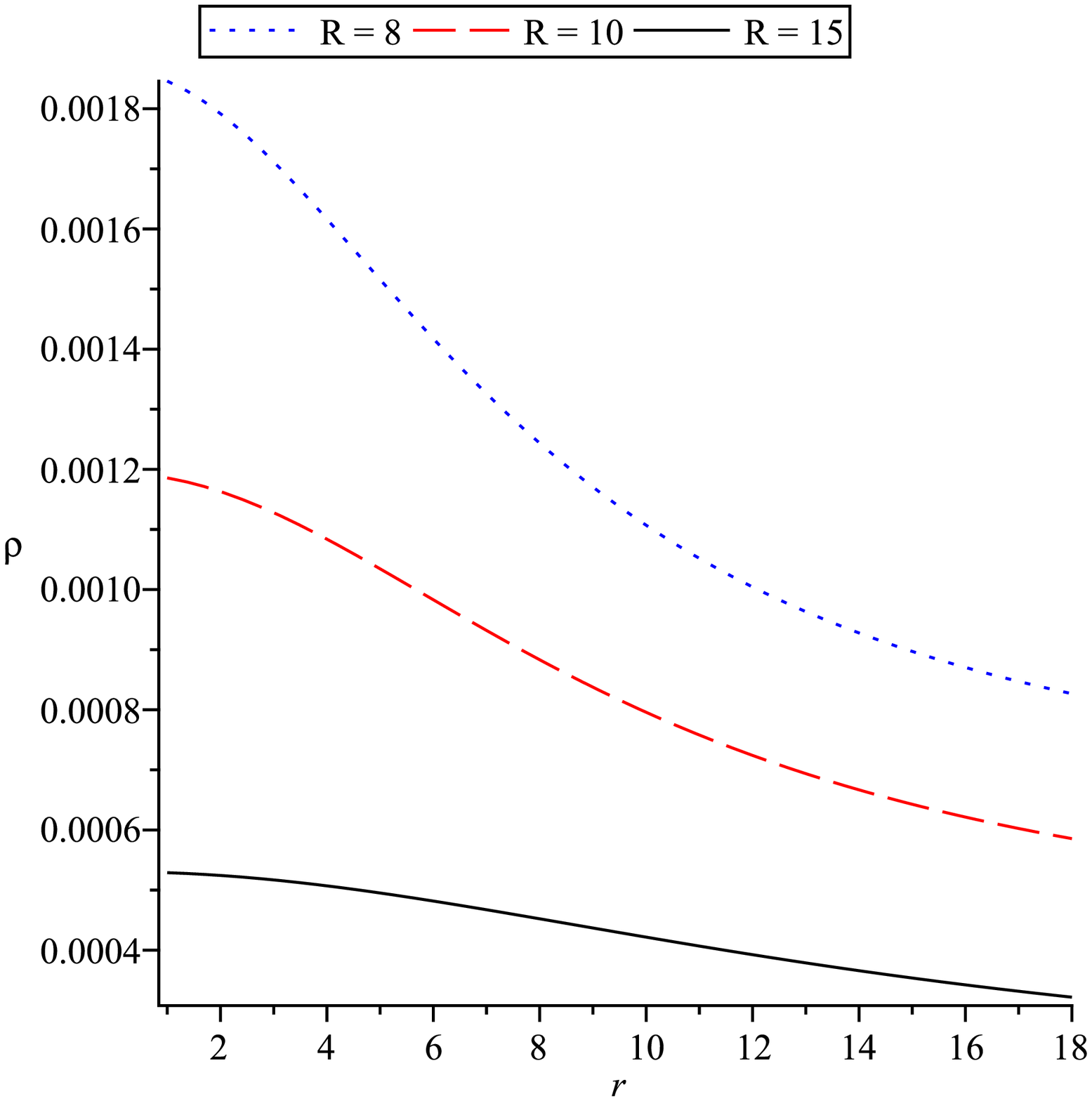}\\
\end{tabular}
\caption{\small{(left) Contour plot for $R$ and $a$ for first boundary condition. (middle) Contour plot for radius and $R$ for third boundary condition. (right) Plot of $\rho$ against $r$.}}
 \end{figure*}

\section{Discussion of the Results}
\subsection{Conformal symmetry}
We have from the conformal equations that
\begin{equation}
e^{\lambda} = \left(\frac{C_{3}}{\psi} \right)^2.
\end{equation}
Thus
\begin{equation}
\psi^2 = \frac{C_{3}^2}{e^{\lambda}} = \frac{C_{3}^2}{1 + \frac{r^2}{R^2}}.
\end{equation}

Since $C_{3}^2 \geqslant 0$ and $R^2 \geqslant 0$, hence $\psi^2 \geqslant 0$. So the compact star admits conformal symmetry. It is interesting to note that the conformal factor does not show any time dependence although we have considered nonstatic conformal symmetry inside the star.

\begin{figure*}[thbp]
\begin{tabular}{lcr}
\includegraphics[width=5.0cm]{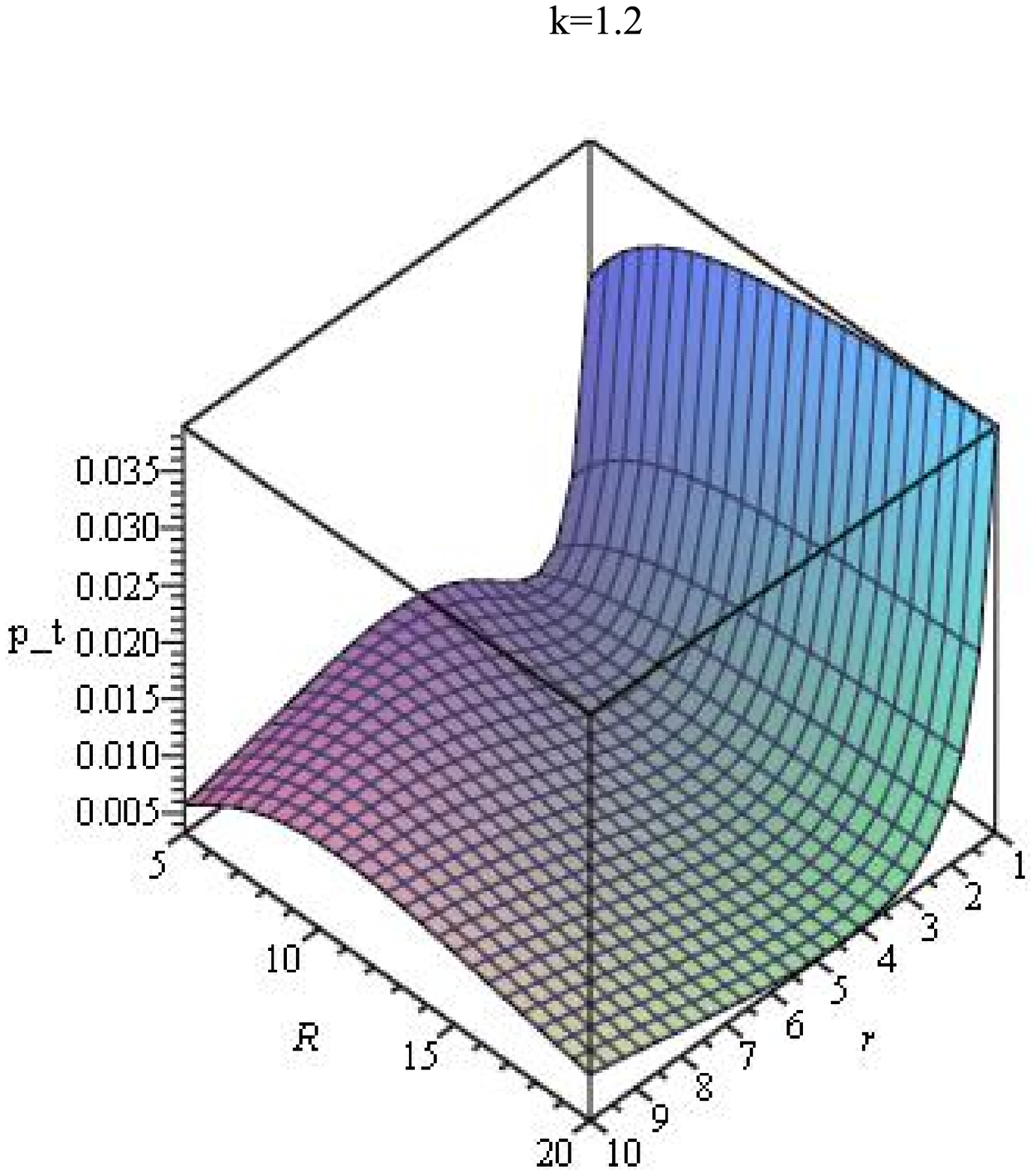}&
\includegraphics[width=5.0cm]{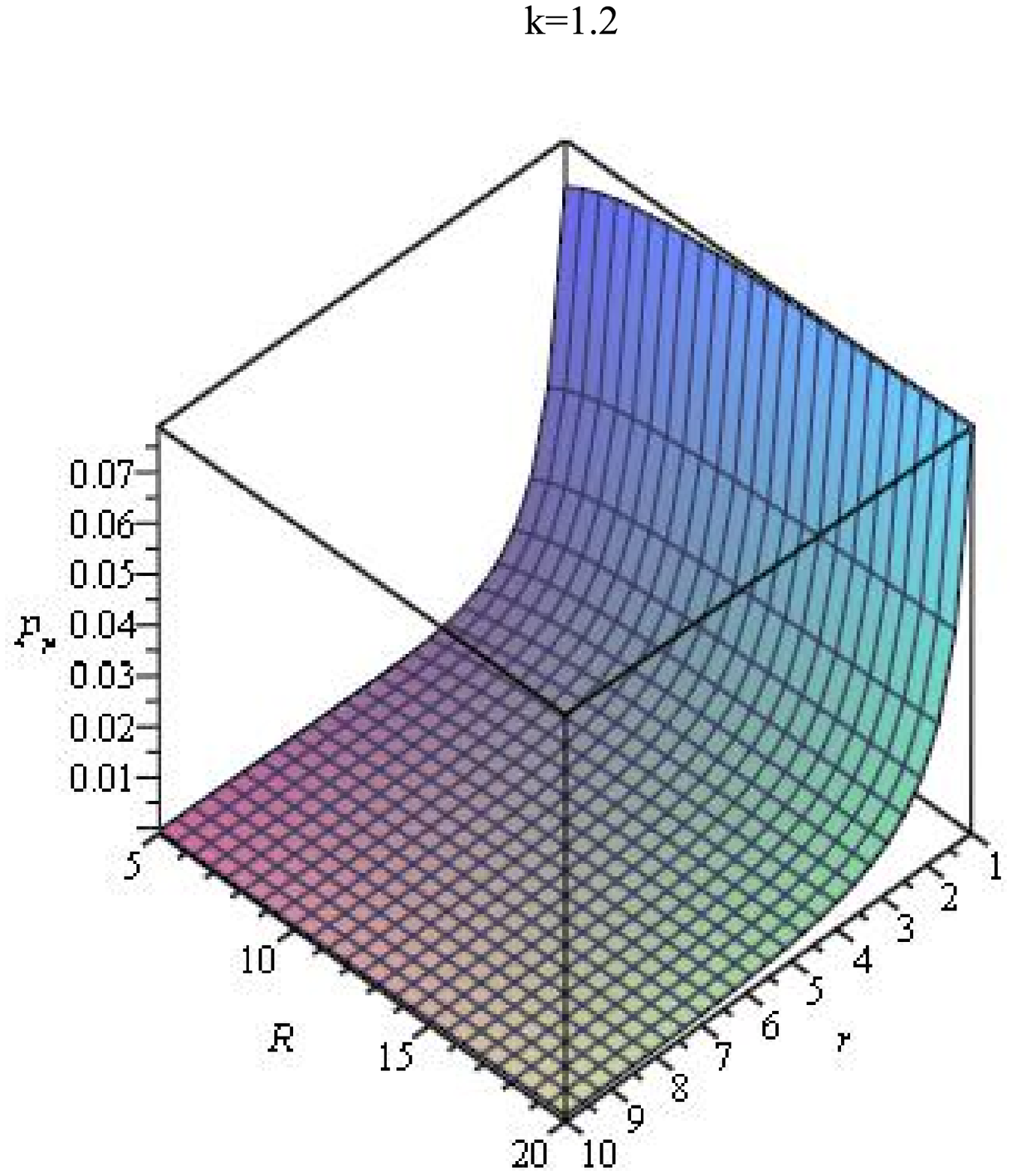}
\includegraphics[width=5.0cm]{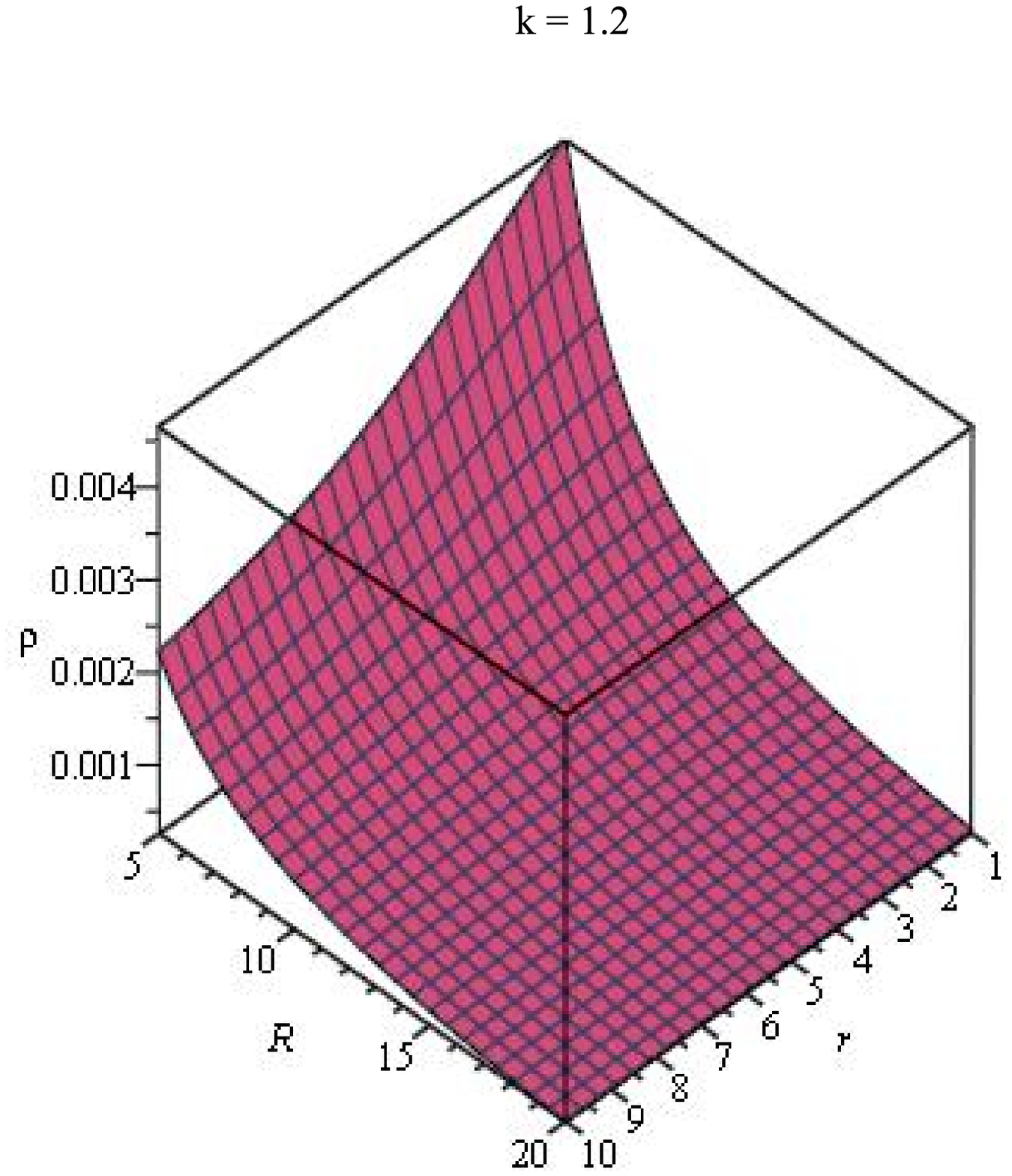}\\
\end{tabular}
\caption{ \small{(Left) Variation of transverse pressure with $r$ and $R$.
        (Middle) Variation of the radial pressure with $r$ and $R$. (Right) Variation of the density with $r$ and $R$.} }
\end{figure*}

\subsection{Density function}
~~~From the three dimensional plot (Figure (1) left panel) of the mass function, it can be noted that the mass of the star depends on the parameter $R$ for a given radial coordinate. As $R$ decreases, the mass of the star increases. As can be seen from equation (\ref{rho}) that the density function is regular at the center of the star. Moreover, from the plot (right panel of Figure (2)) of the density function against radial coordinate, it is obvious that the function is monotonically decreasing with radial distance within the star. Also from right panel of Figure (3) the variation of the density function with the variation of radial coordinate and the parameter $R$ can be noted. So, the behavour of the density function is physically acceptable.

\subsection{Nature of anisotropy}

In the right panel of Fig (1) the variation of anisotropy function ($ \Delta = p_r - p_t $) inside the star is shown for three different choices of $R$. The anisotropy becomes zero at different radial distances for different values of $R$. We can hardly predict any physical feature like anisotropy at the core of the star due to the irregularity of the radial and transverse pressure at the center. However, from the aforementioned figure it can be noted that the anisotropy function is well behaved within the star except at the core region.

\subsection{Radial and transverse pressure}




~~~The transverse pressure decreases continuously within the star upto the boundary. The radial pressure also decreases with radial distance. However, the radial and transverse pressures are arbtrarily large at the center. In our plots we avoided the core region of the star and all plots start from a very small radial distance of the star. The left panel of figure (3) shows the variation of the transverse pressure and middle panel of the same figure shows the variation of radial pressure with radial coordinate and the parameter $R$. The physical analysis for the core region is beyond the scope of the present paper.

\begin{figure*}[thbp]
\begin{tabular}{lr}
\includegraphics[width=5.0cm]{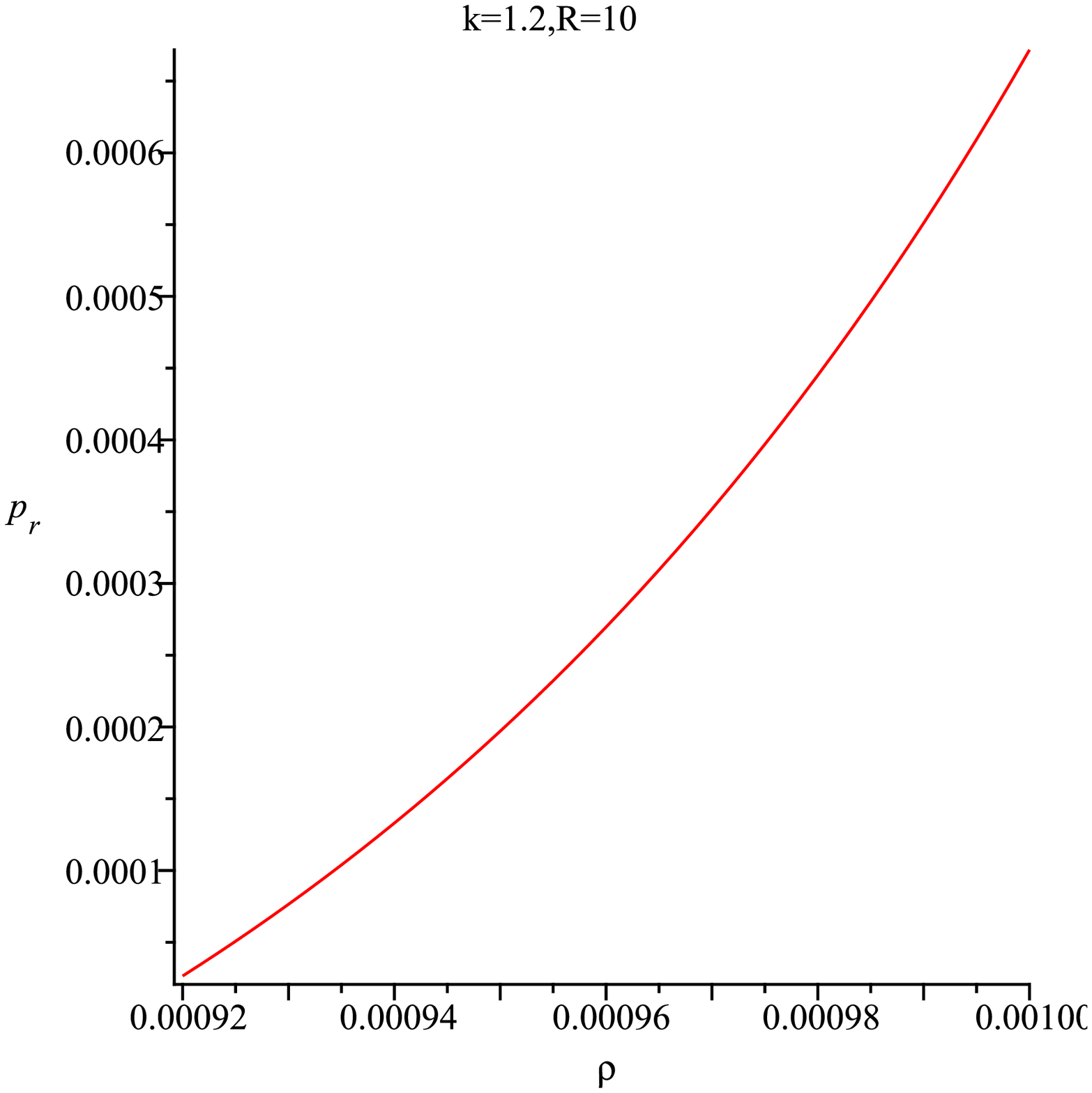}&
\includegraphics[width=5.0cm]{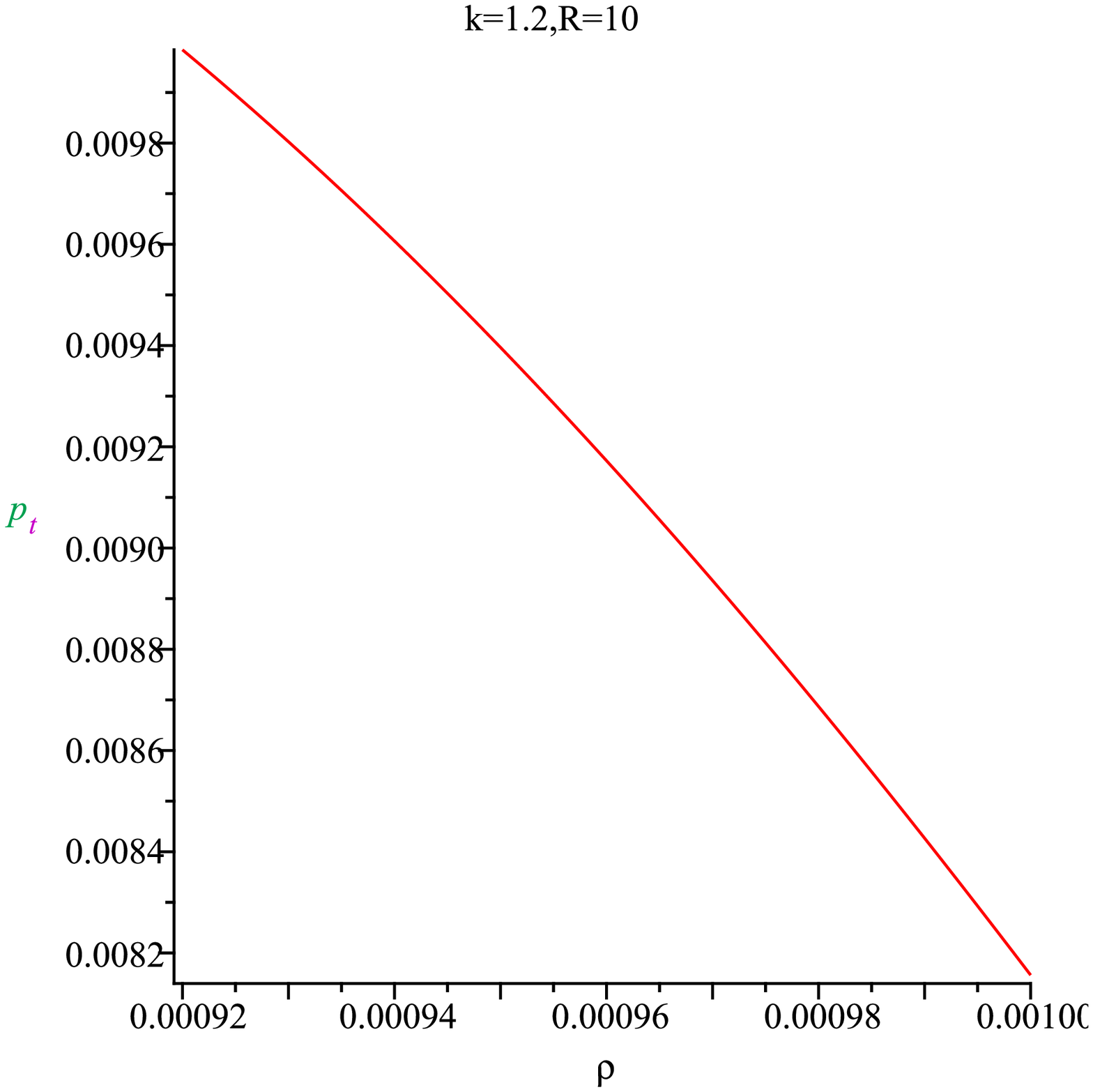}\\
\end{tabular}
\caption{ \small{(Left) The variation of radial pressure $p_r$ against  $\rho$.
        (Right) Variation of the transverse pressure $p_t$ with density.} }
\end{figure*}

\subsection{Equation of state of interior matter}
~~~From the plots (left panel of Figure $4$) for radial pressure against density, we can predict that the relation between the two quantities is nonlinear. Similar behaviour can be noted from the graph of the variation of transverse pressure with density. Thus, the graphs (Figure $4$) imply that the equation of state of the matter inside the star must be nonlinear in nature in both the principal directions. The relation between $p_{r}$ and $\rho$ is as follows
\begin{equation}
8\pi p_r  =  \left(\frac{2}{8\pi \rho R^2 -1} \right)^{-1} \left[\frac{3}{R^2\left(\frac{2}{8\pi \rho R^2-1}-1\right)}-\frac{2K}{\sqrt{\frac{2}{8\pi \rho R^2 -1}}} \left(\frac{1}{R^2} + \frac{1}{R^2 \left(\frac{2}{8\pi \rho R^2-1}+1\right)}\right)\right]-\frac{1}{R^2\left(\frac{2}{8\pi \rho R^2 -1}-1\right)}.
\end{equation}

The relation between $p_t$ and $\rho$ is given by
\begin{eqnarray}
8{\pi}{p_t} & = & \left(\frac{2}{8\pi\rho{R^2}-1}\right)^{-1} \Bigg[\frac{2K\left(\frac{2}{8\pi\rho{R^2}-1}-1\right)}{\sqrt{\frac{2}{8\pi\rho{R^2}-1}}
\left(1+\frac{2}{8\pi\rho{R^2}-1}\right)}-\frac{K}{\sqrt{\frac{2}{8\pi\rho{R^2}-1}}}\left(\frac{1}{R^2}+
\frac{1}{R^2\left(1+\frac{2}{8\pi\rho{R^2}-1}\right)}\right)\left(\frac{1+8\pi\rho{R^2}}{2}\right)\nonumber \\
 &   & \mbox{}   +\left\{\frac{1}{R\sqrt{\frac{2}{8\pi\rho{R^2}-1}-1}}-\frac{KR\sqrt{\frac{2}{8\pi\rho{R^2}-1}-1}}{\sqrt{\frac{2}{8\pi\rho{R^2}-1}}} \left(\frac{1}{R^2}+\frac{1}{R^2\left(1+\frac{2}{8\pi\rho{R^2}-1}\right)}\right)\right\}^{2}\Bigg]\nonumber\\
&  & \mbox{}			 -\frac{1}{R^2}\left(\frac{2}{8\pi\rho{R^2}-1}\right)^{-2}\left\{2-\frac{K{R^2}\left(\frac{2}{8\pi\rho{R^2}-1}-1\right)}{\sqrt{\frac{2}{8\pi\rho{R^2}-1}}} \left(\frac{1}{R^2}+ \frac{1}{R^2{\left(\frac{2}{8\pi\rho{R^2}-1}+1\right)}}\right)\right\}.
\end{eqnarray}
Recently, Ngubelanga et al \cite{ngubelanga} proposed analytical solutions for strange star, considering the quadratic equation of state of the form $p_r = \eta \rho^2 + \alpha \rho - \beta$, where $\eta$, $\alpha$, and $\beta$ are arbitrary constants. The above equations clearly show even more complexity and nonlinearity of the equation of state in both transverse and radial directions.

\subsection{Kretschmann scalar and singularity at $r = 0$}
In the General Theory of Relativity, the Kretschmann scalar is an important quantity to study the physical singularity at a point. It  is defined as
\begin{equation}
R^{\alpha \beta \gamma \delta} R_{\alpha \beta \gamma \delta} = K_{r},
\end{equation}
where $R_{\alpha \beta \gamma \delta}$ is the Riemann tensor. For equation (\ref{metric}) the Kretschmann scalar, $K_{r}$,  is given by
\begin{eqnarray}
K_{r} & = & \frac{1}{4(R^2 + r^2) r^2} \Bigg[ 48 R^4 r^2 + 12 R^4 r^4 (\nu^\prime)^2 + 8 R^6 r^4 \nu^{\prime\prime} (\nu^\prime)^2 - 8 R^6 r^3 \nu^{\prime\prime} \nu^\prime  \nonumber \\
&  & \mbox{} + 8 R^6 r^4 (\nu^{\prime\prime})^2 + 4 R^8 r^2 \nu^{\prime\prime} (\nu^\prime)^2 + 4 R^4 r^6 \nu^{\prime\prime}(\nu^\prime)^2 - 8 R^4 r^5 \nu^{\prime\prime} \nu^\prime + 2 R^6 r^4 (\nu^\prime)^4 - 4 R^6 r^3 (\nu^\prime)^3 + 4 R^8 r^2 (\nu^{\prime\prime})^2 \nonumber \\
&  & \mbox{}  + 4 R^4 r^6 (\nu^{\prime\prime})^2 + R^8 r^2 (\nu^\prime)^4 - 4 R^4 r^5 (\nu^\prime)^3 + 32 R^2 r^4 + 16 r^6 + 8 R^8 (\nu^\prime)^2 + 16 R^6 r^2 (\nu^\prime)^2 \Bigg],
\end{eqnarray}
where
\begin{equation}
\nu^\prime = \frac{2}{r} - \frac{2 K r}{\left(\sqrt{1 + \frac{r^2}{R^2}}\right) R^2} + \frac{r}{\left(1 + \frac{r^2}{R^2} \right)^{\frac{3}{2}} R^2 \left(1 + \frac{1}{1 + \frac{r^2}{R^2}} \right)}
\end{equation} \\
and
\begin{eqnarray}
\nu^{\prime \prime} & = & - \frac{2}{r^2} + \frac{2 K r^2}{\left(1 + \frac{r^2}{R^2} \right)^{\frac{3}{2}} R^4} - \frac{2 K}{\left(\sqrt{1 + \frac{r^2}{R^2}}\right) R^2} - \frac{3 r^2}{\left(1 + \frac{r^2}{R^2} \right)^{\frac{3}{2}} R^4 \left(1 + \frac{1}{1 + \frac{r^2}{R^2}} \right)} + \frac{1}{\left(1 + \frac{r^2}{R^2} \right)^{\frac{3}{2}} R^2 \left(1 + \frac{1}{1 + \frac{r^2}{R^2}} \right)} \nonumber \\
& & \mbox{} + \frac{2 r^2}{\left(1 + \frac{r^2}{R^2} \right)^{\frac{7}{2}} R^4 \left(1 + \frac{1}{1 + \frac{r^2}{R^2}} \right)^2}.
\end{eqnarray}
From the expression for the Kretschmann scalar, it is apparent that there is point of divergence at $r = 0$, i.e., the spacetime possesses a singularity at the origin. Apart from this physical singularity our solution is regular everywhere.

\section{Tests of the Model:}
\subsection{Compactness parameter:}
The compactness parameter is defined as $u= \frac{ m(r)}{r}$  and for the present model it is given by

\begin{equation}
u(r) = \frac{\int_{0}^{r}4 \pi r^2 \left( \frac{1}{8 \pi R^2} \right) \left( 1 - \frac{r^2}{R^2}\right)^{-1} dr}{r} = \frac{r^2}{600} + 1 - \frac{10 \arctan{\left(\frac{r}{R}\right)}}{r}.
\end{equation}
Buchdahl \cite{buchdahl} showed that for a
perfect fluid sphere  $2M/R\leqslant 8/9$ where $M$ is the maximum allowed
mass and $R$ is the radius. Moreover,
if the trace of the energy momentum tensor is postulated
to be nonnegative, then the ratio of the total mass to the
coordinate radius is $\leqslant 5/18$.  In other words, $M/R$ is
strictly less than 4/9.  Now, while we would expect the
quantity $\left(1-2M/R\right)$ to be nonnegative, Buchdahl's condition
actually does not allow the value to be less than $\frac{1}{9}$. Adler \cite{adler} found from closed analytic solution of Einstein's equation that for static fluid sphere the maximum mass is $\frac{2}{5}^{th}$ of the radius in geometric units. The result obtained by Adler is consistent with the result of Buchdahl. From the three dimensional plots it can be noted that the radial coordinate up to which Buchdahl's condition is satisfied inside the star is dependent on the value of $R$. For $R \leqslant 7$ the aforementioned conditions are violated for smaller values of radial coordinate. As can be seen from table II, the smaller values of the parameter $R$ correspond to smaller possible radii of the star. Thus the observations from the graph are consistent with the results of table II.

~~~~ From figure 5 it can be noted that the value of the compactness parameter is safely below the upper limit set by Buchdahl for $R \geqslant 10$ up to the stellar radius of $10$ km.


\subsection{Surface redshift for the star:}

\begin{figure*}[thbp]
    \begin{tabular}{lr}
    \includegraphics[scale=0.3]{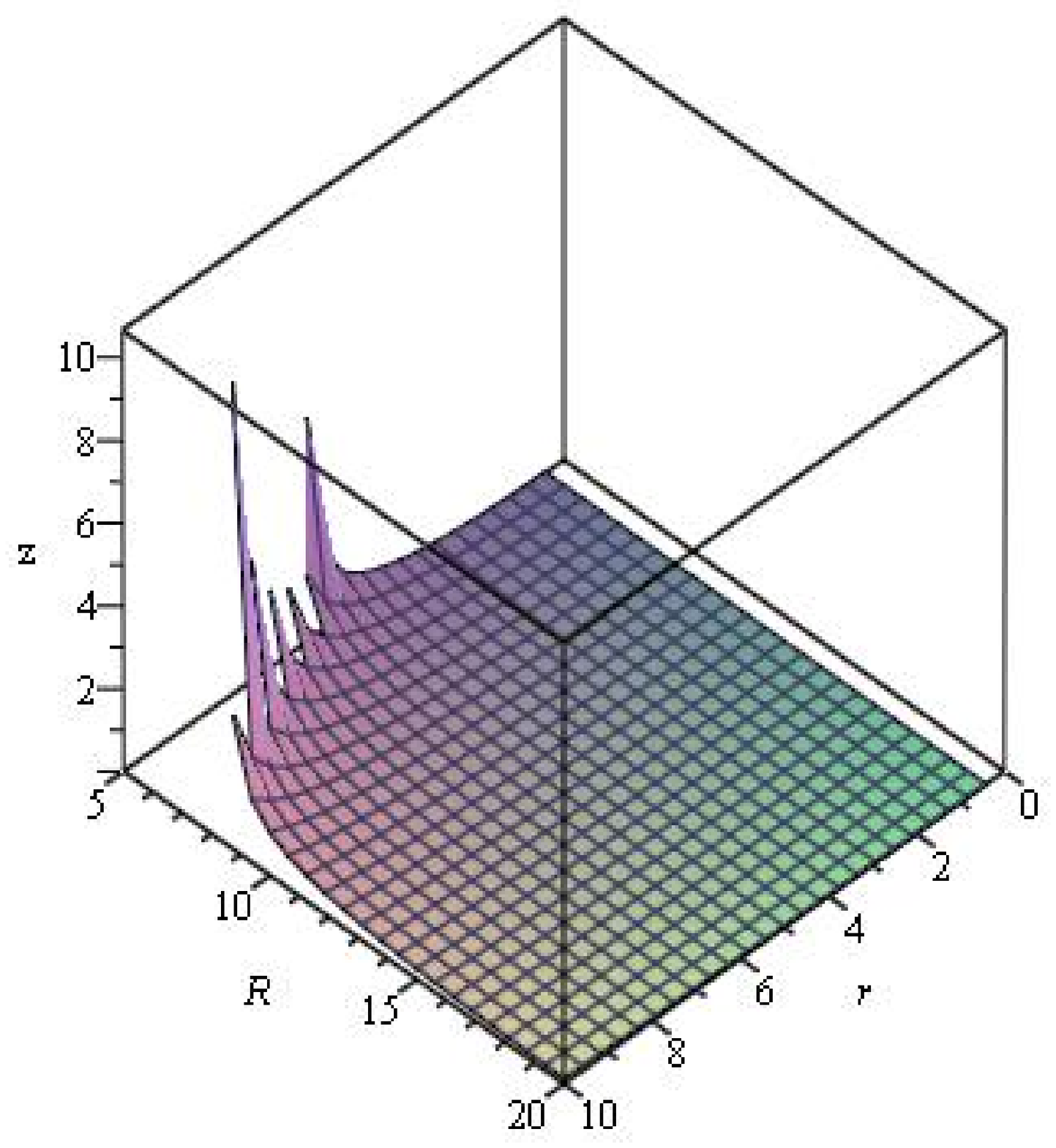}&
    \includegraphics[scale=0.3]{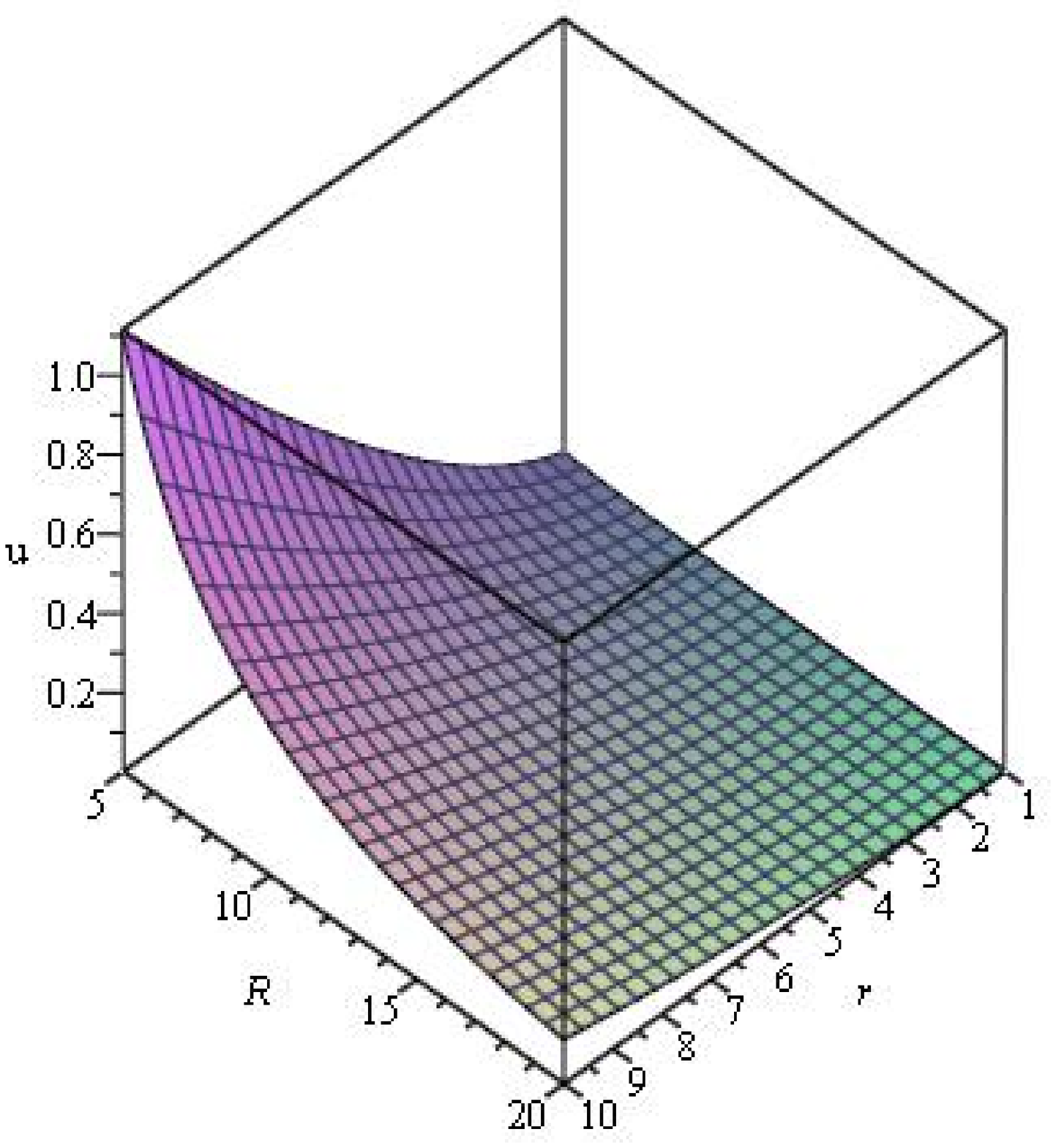} \\
    \end{tabular}
    \caption{\small{(left) Variation of redshift function with radial distance and $R$. (right)Variation of compactness against radial distance and $R$}}
\end{figure*}

The surface redshift $Z$ corresponding to the compactness $u$ is given by
\begin{equation}
1+Z= \left[ 1-(2 u )\right]^{-\frac{1}{2}}. \label{z}
\end{equation}
The redshift $Z$, which is an indirect measure of the compactness of the star, can be measured from the X-ray spectrum. High observed
redshifts are consistent  with strange stars which have mass-radius ratios higher than neutron stars. From figure (5), we note that the for smaller values of the parameter $R$, the redshift of the star seems to be unphysical. As the parameter value increases, the surface red shift of the star decreases. However, one can also study the two dimensional graphs of the function for three chosen values of $R$ in figure 6.The values of the redshift function remain physical for $R = 10, 15, 20$.
As can be seen from the figure 6 the redshift values are very high. That high values of the surface redshift are consistent with the strange stars, was noted by Rahaman et al. \cite{farook}.

\begin{figure*}[thbp]
    \begin{tabular}{ll}
    \includegraphics[scale=0.3]{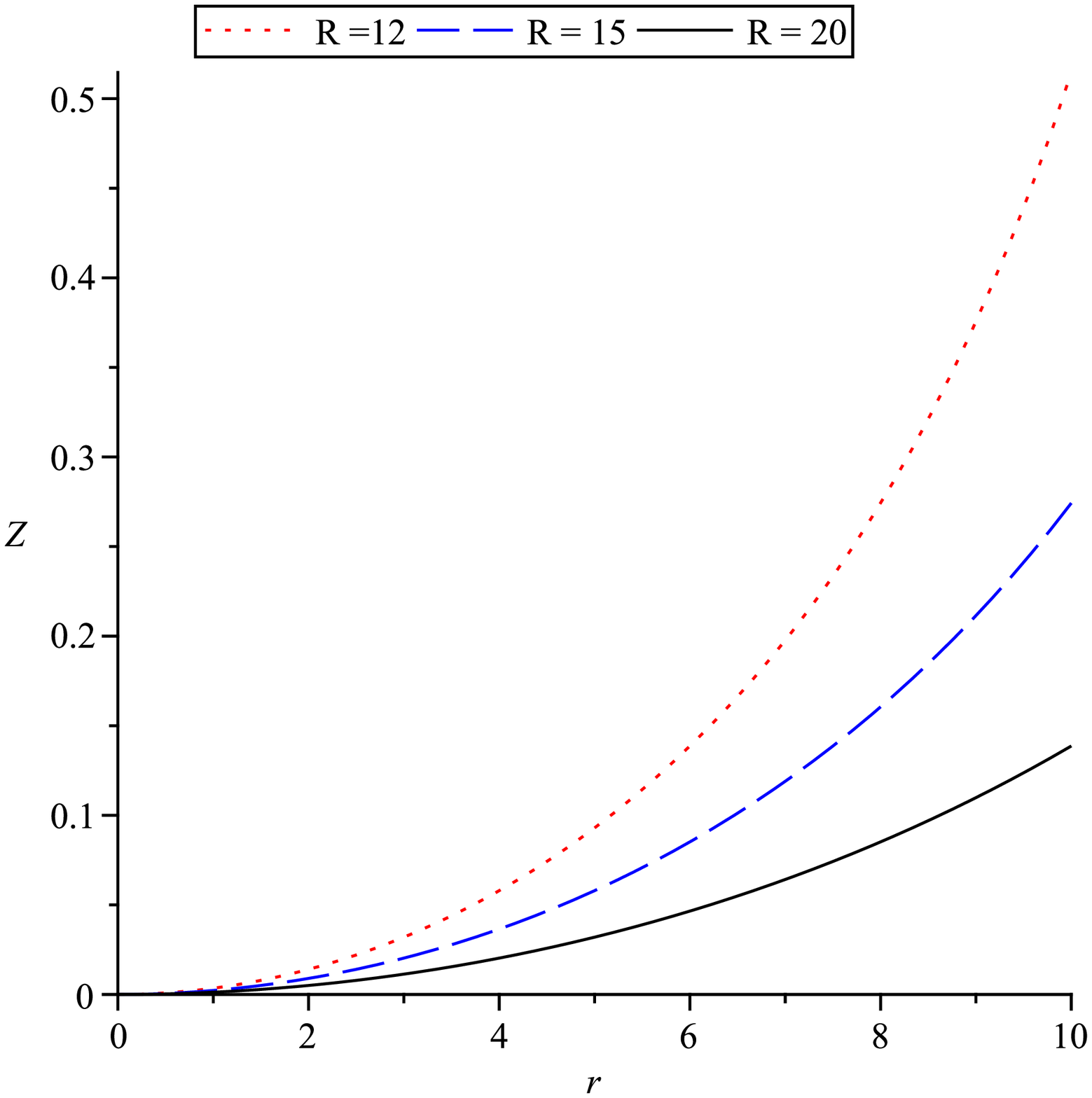}&
    \includegraphics[scale=0.3]{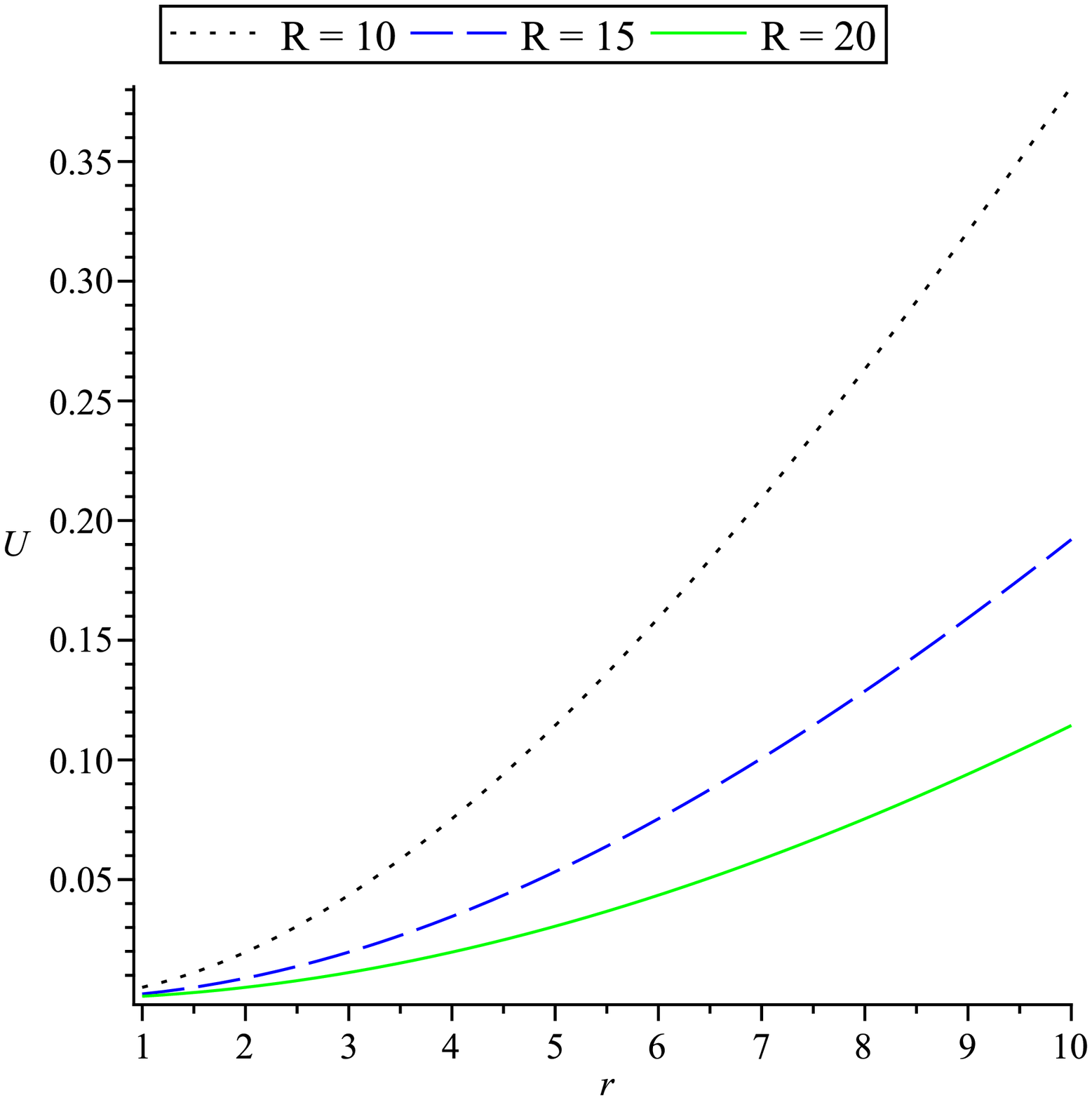} \\
    \end{tabular}
    \caption{\small{(left) Variation of redshift function with radial distance for different $R$. (right)Variation of compactness against radial distance with different $R$}.}

\end{figure*}

\subsection{Energy conditions:}
The four energy conditions, namely the null, weak, strong and dominant, are important for the physical analysis of any gravitational system. For the first three energy conditions to be satisfied, are following inequalities are to be satisfied within the star
\begin{eqnarray}
\rho \geqslant 0, \\
p_r + \rho \geqslant 0, \\
p_t + \rho \geqslant 0, \\
p_r + 2p_t + \rho \geqslant 0.
\end{eqnarray}

As can be noted from the Figure $7$, the null, weak and strong energy conditions are satisfied everywhere within the star except at the center. Since the solutions for $p_r$ and $p_t$ are not well behaved near the center of the star, we cannot speak about any physical features in that region.
The energy density is positive, however, everywhere within the star.
\begin{figure}[thbp]
\begin{tabular}{ll}
\includegraphics[width=5.0cm]{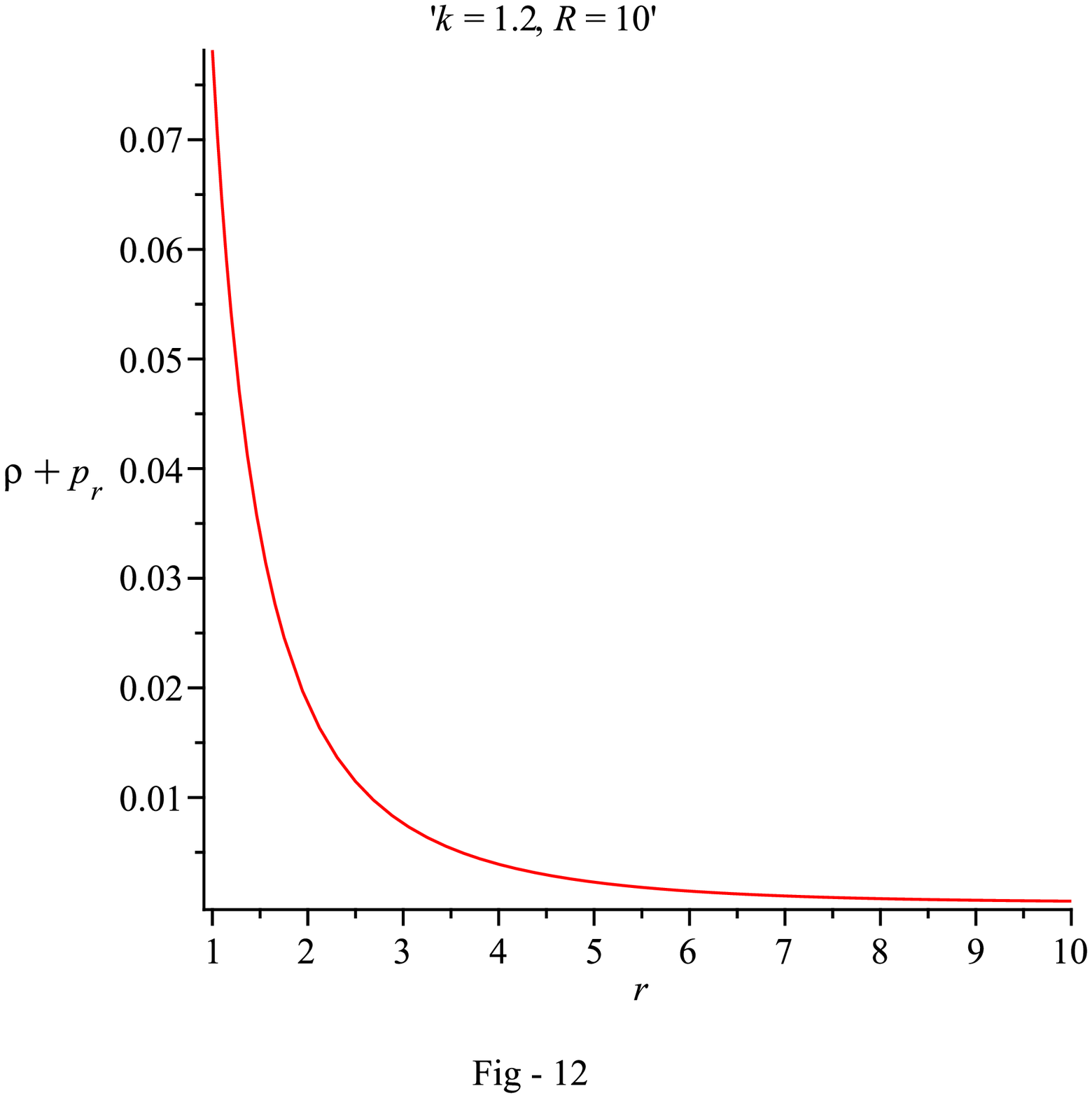}&
\includegraphics[width=5.0cm]{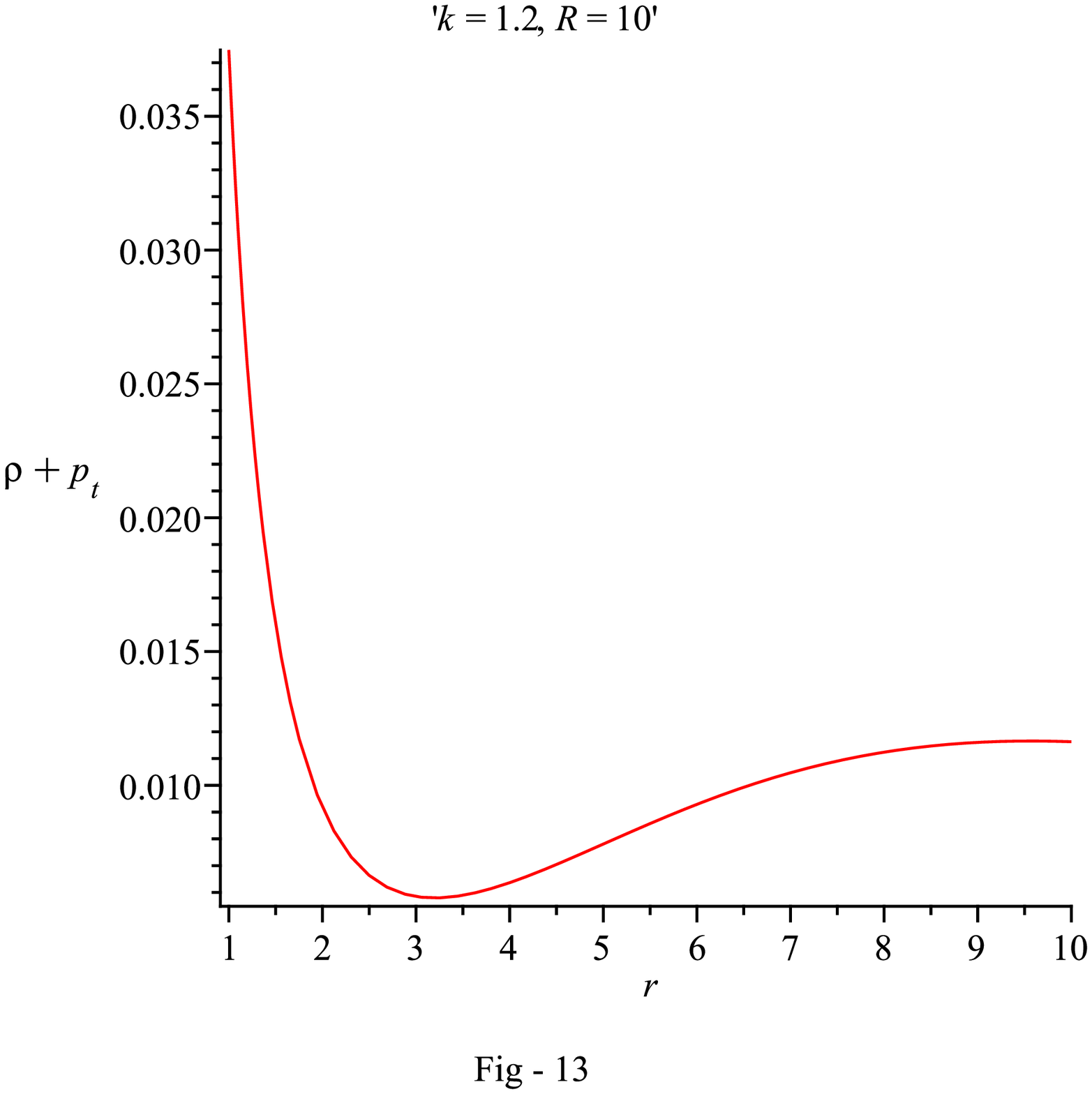}
\includegraphics[width=5.0cm]{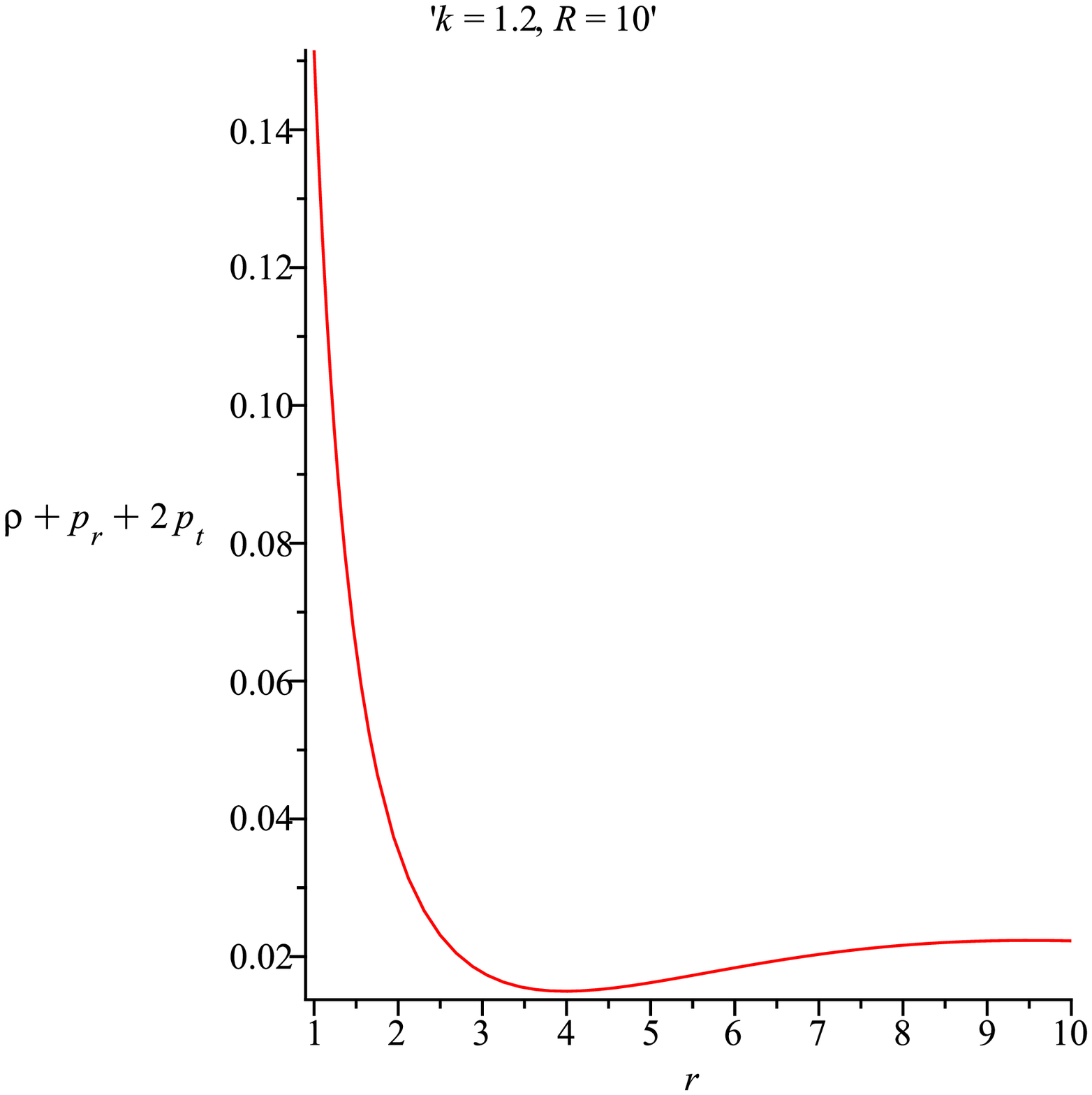}\\
\end{tabular}
\caption{ \small{Null, weak and strong energy conditions within the star}. }
\end{figure}

\begin{table*}
\centering \caption{The predicted values of the mass and radius for  various
compact stars form various observations}

\begin{tabular}{|c|c|c|c|} \hline
Compact Stars & Radius ( in km)   & Mass ($M_{\odot}$) &  Mass in km \\
\hline PSR J1614-2230 & $10.3$
 & $1.97 \pm 0.04$ & $2.9057 \pm 0.059$ \\
\hline Vela X - 12 & $9.99$ & $1.77 \pm 0.08$ & $2.6107 \pm 0.118$ \\
\hline  PSR J1903+327 & $9.82$
 & $1.667 \pm 0.021$ & $2.4588 \pm 0.03$ \\ \hline Cen X - 3  & $9.51$ & $1.49 \pm 0.08$ & $2.1977 \pm 0.118$  \\
 \hline SMC X - 1 & $9.13$ & $1.29 \pm 0.05$ & $1.9027 \pm 0.073$ \\
  \hline
  \end{tabular}
\end{table*}





\begin{table*}
{\small \centering \caption{Theoretical values of mass of the compact objects for diffrent choices of $R$. The radii are calculated from the Boundary condition $p_r = 0$ and $K = 1.2$}

\begin{tabular}{|c|c|c|c|c|c|} \hline
$R$ ( in km)   & $K$ & \small{$r$(in km) for $p_r = 0$} & Mass(km) & Mass $(M_{\odot})$ & Remarks\\
\hline 10 & 1.2 & 7.560685876 & 1.807280206 & 1.225 $(M_{\odot})$ & \\
\hline 10.1 & 1.2 & 7.636292735 & 1.904713826 & 1.29 $(M_{\odot})$ & $\sim$ SMC X - 1 \\
\hline 10.2 & 1.2 & 7.711899594 & 2.002583967 & 1.3577 $(M_{\odot})$ &  \\
\hline 10.3 & 1.2 & 7.787506452 & 2.100894950 & 1.42 $(M_{\odot})$ & \\
\hline 10.4 & 1.2 & 7.7863113311 & 2.146492574 & 1.455 $(M_{\odot})$ & $\sim$Cen X - 1 \\
\hline 10.41 & 1.2 & 7.870673997 & 2.209551364 & 1.5 $(M_{\odot})$ & $\sim$ Cen X - 1 \\
\hline 10.5 & 1.2 & 7.938720170 & 2.298720170 & 1.558 $(M_{\odot})$ &  \\
\hline 10.6 & 1.2 & 8.014327029 & 2.398516178 & 1.626 $(M_{\odot})$ & \\
\hline 10.61 & 1.2 & 8.021887715 & 2.408507249 & 1.633 $(M_{\odot})$ & \\
\hline 10.62 & 1.2 & 8.029448400 & 2.418502903 & 1.6396 $(M_{\odot})$ & \\
\hline 10.63 & 1.2 & 8.037009086 & 2.428503149 & 1.646 $(M_{\odot})$ & \\
\hline 10.64 & 1.2 & 8.044569772 & 2.438507989 & 1.653 $(M_{\odot})$ & \\
\hline 10.66 & 1.2 & 8.059691144 & 2.458531470 & 1.666 $(M_{\odot})$ & $\sim$ PSR J1930 + 327 \\
\hline 10.7 & 1.2 & 8.089933887 & 2.498633752 & 1.69 $(M_{\odot})$ & \\
\hline 10.8 & 1.2 & 8.165540746 & 2.599213779 & 1.76 $(M_{\odot})$ & \\
\hline 10.81 & 1.2 & 8.173101432 & 2.609297383 & 1.769 $(M_{\odot})$ & \\
\hline 10.82 & 1.2 & 8.180662118 & 2.619385660 & 1.77 $(M_{\odot})$ & $\sim$ Vela X - 1 \\
\hline 10.9 & 1.2 & 8.241147605 & 2.700260582 & 1.83 $(M_{\odot})$ & \\
\hline 11 & 1.2 & 8.316754464 & 2.801778481 & 1.899 $(M_{\odot})$ & \\
\hline 11.1 & 1.2 & 8.392361323 & 2.903771799 & 1.9686 $(M_{\odot})$ & $\sim$ PSR J1614 - 2230 \\

  \hline\end{tabular} }
\end{table*}

\section{Concluding Remarks}
In the foregoing, we have presented a class of solutions of Einstein's field equations for a compact spherically symmetric star with anisotropic principal pressures. The compact object is assumed to admit nonstatic conformal symmetry. It is observed that the star admits conformal symmetry, but the solutions of the field equations are static. It seems that since the background spacetime is assumed to be static, the time dependence of the conformal symmetry is not \textquotedblleft seen \textquotedblright by the field equations.
We have shown that the null, weak and strong energy conditions of general relativity are satisfied within the star. This confirms that the matter content inside the star is real matter. On the other hand, from the plot of principal pressures against the density, as well as from their analytic expressions, one can note that the equation of state of the matter inside the star has a complicated nonlinear form. Apart from this, the present model also satisfies the Buchdahl limit for the quantity $\frac{2 M}{r}$. The surface redshift function also gives physical values for different values of the parameter $R$.
From the study of the Kretschman curvature scalar, we show that the present model has a physical singularity at the center of the star. Except for this feature, the solution is regular everywhere.
Some interesting features of our solution can be noted from Table II. For different choices for the parameter $R$, we get the solutions for compact stars of masses nearly equal to the observed stars like SMC X - $1$, Cen X - $3$, PSR J$1903+327$, Vela X - $1$ and PSR J$1614 - 2230$. The radii, obtained from the present model, are also nearly equal to the observationally predicted radii. We predict smaller values for radii of the aforementioned observed compact stars \cite{farook}. But considering the fact that the determination of the radii of these compact objects suffers from systematic errors, we can surely ignore the mismatch between the radii predicted from the present model and that from observational predictions.

\section*{Acknowledgments} FR and KC are thankful to the Inter-University Centre for Astronomy and Astrophysics, Pune,
India for providing them Visiting Associateships under which a part of this work was carried out. KC is thankful to the University Grants Commision for providing financial support in Minor Research Project under which this research work was carried out. SDM further acknowledges that this work is supported by the South African Research Chair Initiative of the Department of Science and Technology and the National Research Foundation of South Africa. FR is also grateful to DST-SERB and DST-PURSE,  Govt. of India for financial support. IHS is
also thankful to DST, Government of India, for providing financial
support under INSPIRE Fellowship.

\end{document}